\documentclass[10pt]{article}
\usepackage{graphicx}

\def\Title#1{\begin{center} {\Large #1 } \end{center}}
\def\Author#1{\begin{center}{ \sc #1} \end{center}}
\def\Address#1{\begin{center}{ \it #1} \end{center}}

\newcommand\pubblock{\rightline{\begin{tabular}{l} Proceedings of the Fifth Annual LHCP\\ \pubnumber\\
         \pubdate  \end{tabular}}}

\newenvironment{Abstract}{\begin{quotation} \begin{center} 
             \large ABSTRACT \end{center}\bigskip 
      \begin{center}\begin{large}}{\end{large}\end{center} \end{quotation}}

\newenvironment{Presented}{\begin{quotation} \begin{center} 
             PRESENTED AT\end{center}\bigskip 
      \begin{center}\begin{large}}{\end{large}\end{center} \end{quotation}}

\def\Acknowledgements{\bigskip  \bigskip \begin{center} \begin{large}
             \bf ACKNOWLEDGEMENTS \end{large}\end{center}}

\def\beq{\begin{equation}}
\def\eeq#1{\label{#1}\end{equation}}
\def\eeqn{\end{equation}}

\def\beqa{\begin{eqnarray}}
\def\eeqa#1{\label{#1}\end{eqnarray}}
\def\eeqan{\end{eqnarray}}

\let\bar=\overbar

\def\Dslash{\not{\hbox{\kern-4pt $D$}}}
\def\dslash{\not{\hbox{\kern-2pt $\del$}}}

\def\msb{{\bar{\ssstyle M \kern -1pt S}}}

\usepackage{color}

\newcommand\pubnumber{ CMS CR-2017/217 }

\newcommand\pubdate{\today}

\def\affiliation{
On behalf of the CMS Experiment, \\
Vinca Institute of Nuclear Sciences \\
University of Belgrade, Serbia}

\begin{document}

\large
\begin{titlepage}
\pubblock

\vfill
\Title{  ELECTROWEAK PRECISION MEASUREMENTS IN CMS  }
\vfill

\Author{ MILOS DORDEVIC  }
\Address{\affiliation}
\vfill
\begin{Abstract}

An overview of recent results on electroweak precision measurements from the CMS Collaboration is presented. Studies of the weak boson differential transverse momentum spectra, Z boson angular coefficients, forward-backward asymmetry of Drell-Yan lepton pairs and charge asymmetry of W boson production are made in comparison to the state-of-the-art Monte Carlo generators and theoretical predictions. The results show a good agreement with the Standard Model. As a proof of principle for future W mass measurements, a W-like analysis of the Z boson mass is performed.
 
\end{Abstract}
\vfill

\begin{Presented}
The Fifth Annual Conference\\
 on Large Hadron Collider Physics \\
Shanghai Jiao Tong University, Shanghai, China\\ 
May 15-20, 2017
\end{Presented}
\vfill
\end{titlepage}
\def\thefootnote{\fnsymbol{footnote}}
\setcounter{footnote}{0}
%

\normalsize 


\section{Introduction}

Precision measurements of the electroweak observables of the Standard Model (SM) provide one of the most stringent test of its consistency. The data recorded at the highest available energies by the experiments such as ATLAS \cite{ref11} and CMS \cite{ref12} allows to tune the Monte Carlo (MC) generators and provide state-of-the-art theory calculations. Any deviation from the SM prediction could be a sign of new physics. Differential measurement of the weak boson transverse momentum spectrum, angular coefficients of the Z boson, forward-backward asymmetry of Drell-Yan lepton pairs, W boson charge asymmetry and W-like analysis of Z boson mass are presented in this overview.

\section{Weak vector boson transverse momentum spectra}

\begin{figure}[htb]
\begin{minipage}{14pc}
\includegraphics[width=16pc]{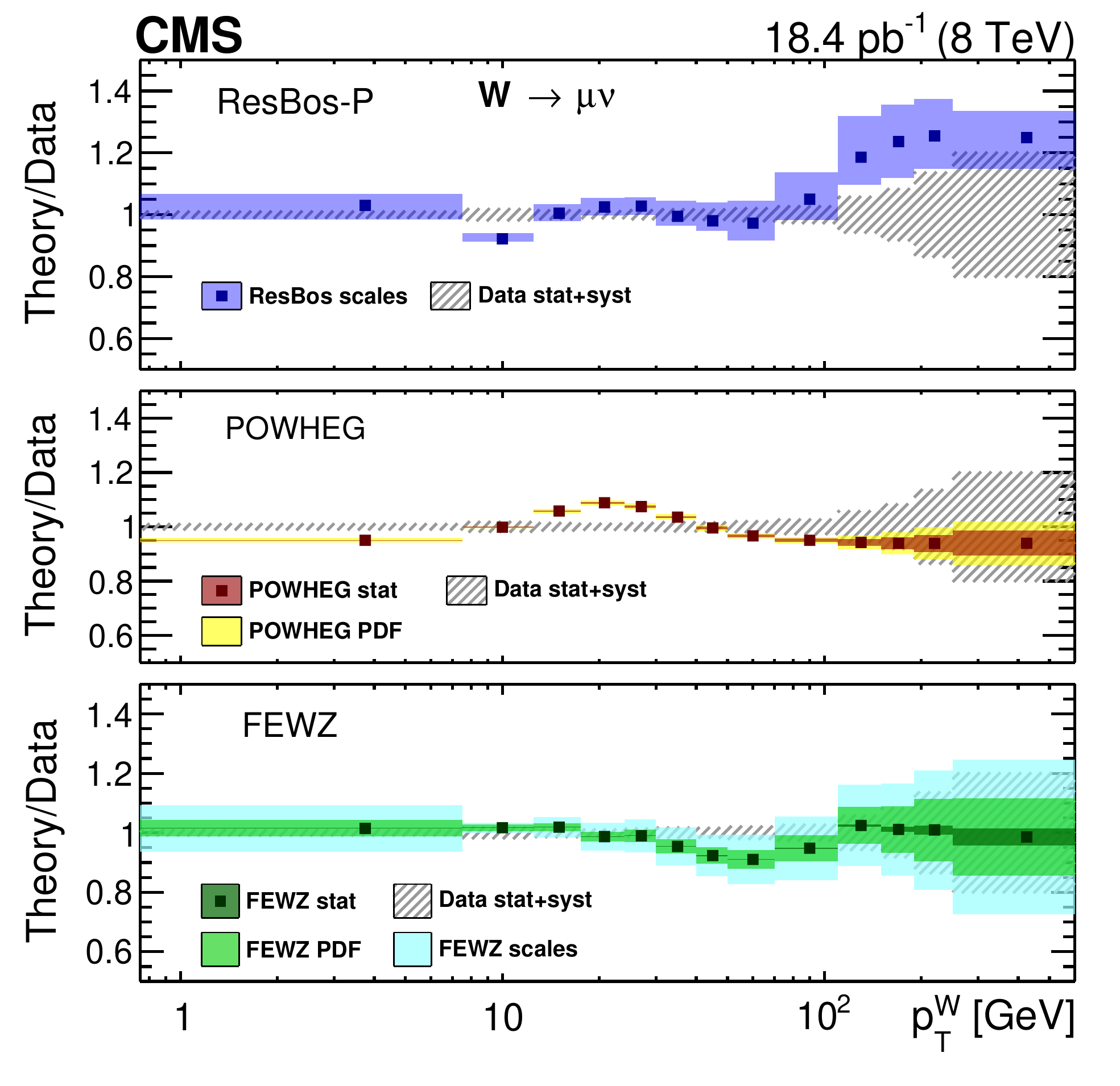}
\caption{\label{Vpt1}Normalized differential cross section for charge independent W boson production as a function of $p_{T}^{W}$ for muon decay \cite{ref21}. }
\end{minipage}\hspace{2pc}%
\begin{minipage}{14pc}
\includegraphics[width=16pc]{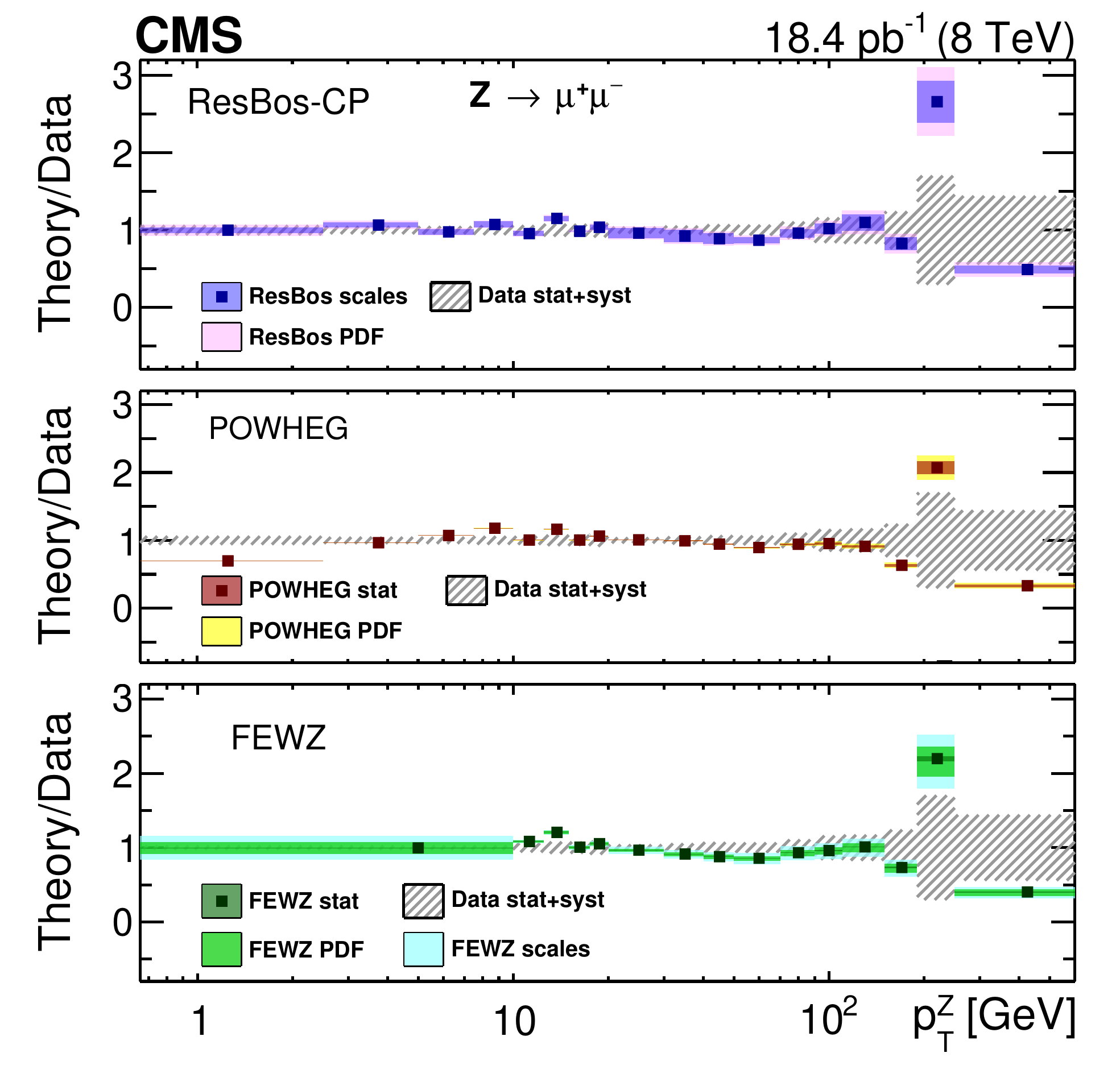}
\caption{\label{Vpt2}Comparison of the normalized dimuon differential transverse momentum distribution with different theoretical predictions \cite{ref21}. }
\end{minipage} 
\end{figure}

The W and Z boson differential cross section was measured \cite{ref21} using a low pile-up run with an average of 4 interactions per bunch crossing. The results were compared to the prediction of RESBOS (P and CP versions) \cite{ref22, ref23, ref24}, POWHEG \cite{ref25} and FEWZ \cite{ref26, ref27, ref28} generators. The ratio of the predicted to the measured differential cross section for the W boson is presented in Fig.~\ref{Vpt1}. In the case of RESBOS-P the agreement is good up to the W boson transverse momentum of 110 GeV, while POWHEG shown an increase of 12$\%$ at 25 GeV. The NNLO FEWZ generator reported a difference of 10$\%$ at the transverse momentum of 60 GeV. The corresponding results for the Z boson shows a good agreement with the RESBOS-CP, the 30$\%$ lower estimate of data in the 0 to 2.5 GeV range and 18$\%$ excess between 7.5 and 10 GeV. In the low transverse momentum region below 20 GeV, the FEWZ generator has a divergent behaviour, as shown in Fig.~\ref{Vpt2}. The $W^{+}/W^{-}$ and $Z/W$ ratios are measured as well and all three generators listed above describe data well within the experimental uncertainties. The ratio of differential Z boson cross section was measured at 7 and 8 TeV in the muon channel and theory prediction provides a good description of data, within the uncertainties.
 
\section{Angular coefficients of Z boson}

\begin{figure}[htb]
\begin{minipage}{14pc}
\includegraphics[width=16pc]{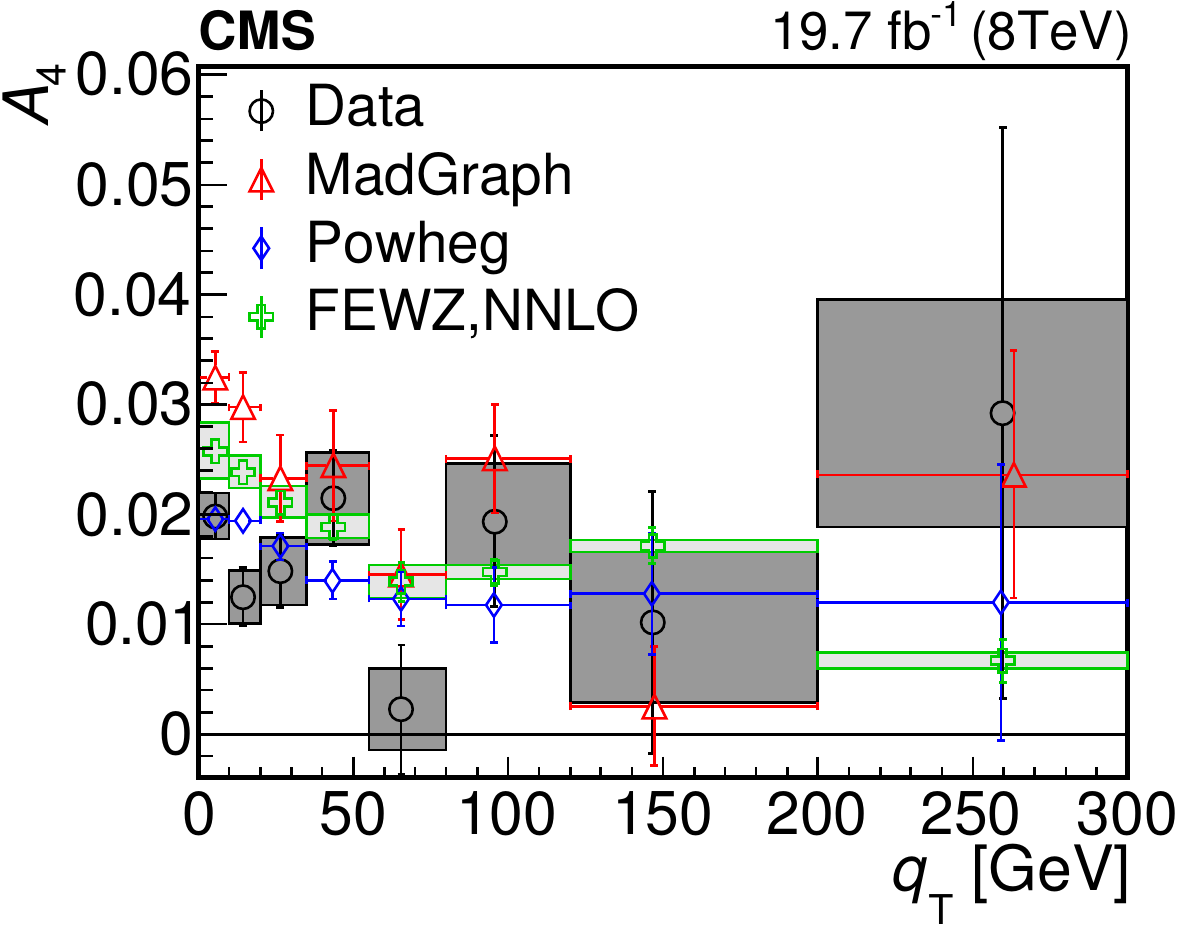}
\caption{\label{AngZ1}The  $A_{4}$ coefficient measured in bins of $q_{t}$ and rapidity up to 1 \cite{ref31}.}
\end{minipage}\hspace{2pc}%
\begin{minipage}{14pc}
\includegraphics[width=16pc]{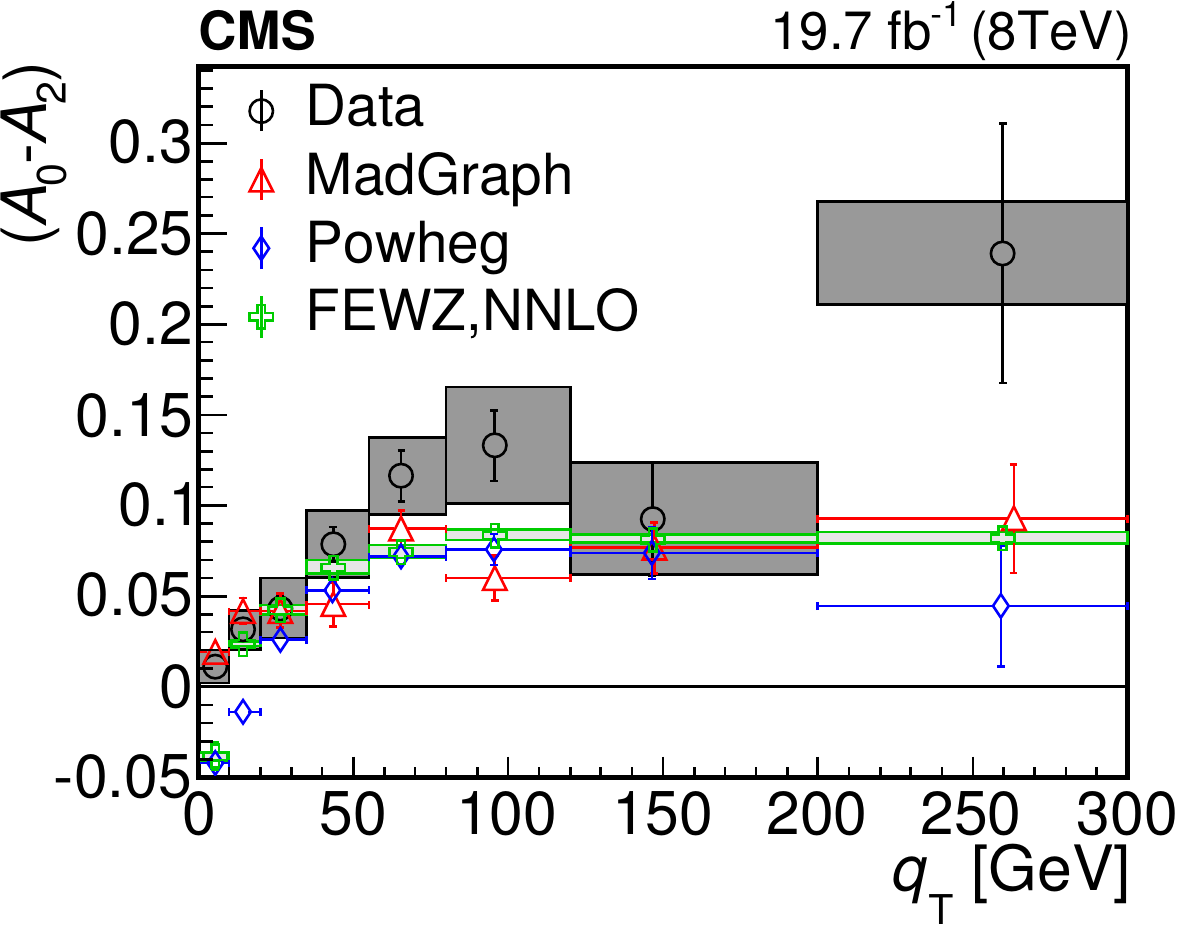}
\caption{\label{AngZ2}The $A_{0}-A_{2}$ measured in bins of $q_{t}$ and rapidity up to 1 \cite{ref31}.}
\end{minipage} 
\end{figure}

The first CMS measurement \cite{ref31} of the angular coefficients of Z boson decaying to muon pair is presented. These coefficients govern the decay of Z boson and thus also the kinematics of the lepton. The values of the coefficients follow from the vector and axial vector structure of boson-fermion couplings. In proton-proton collisions the dominant contribution is the quark annihilation process, while the contribution of the qg Compton process is larger than in the proton-antiproton collisions. The coefficients were measured as a function of the transverse momentum of the two lepton system ($q_{T}$) and in two rapidity bins. The data was compared to the prediction of MadGraph (LO) \cite{ref32}, POWHEG (NLO) and FEWZ (NNLO) generators. The MadGraph generator predicts higher $A_{4}$ coefficient, as shown in Fig.~\ref{AngZ1} as it calculates the $\Theta_{W}$ without considering the radiative corrections. A violation of the Lam-Tung relation is observed in the $A_{0}-A_{2}$ distributions, as shown Fig.~\ref{AngZ2}, in particular for high $q_{T}$ due to missing higher order calculations. The measurement of these coefficients is important for the high-precision measurements of the W mass and electroweak mixing angle.

\section{Forward-backward asymmetry of Drell-Yan lepton pairs}

\begin{figure}[htb]
\begin{minipage}{14pc}
\includegraphics[width=16pc]{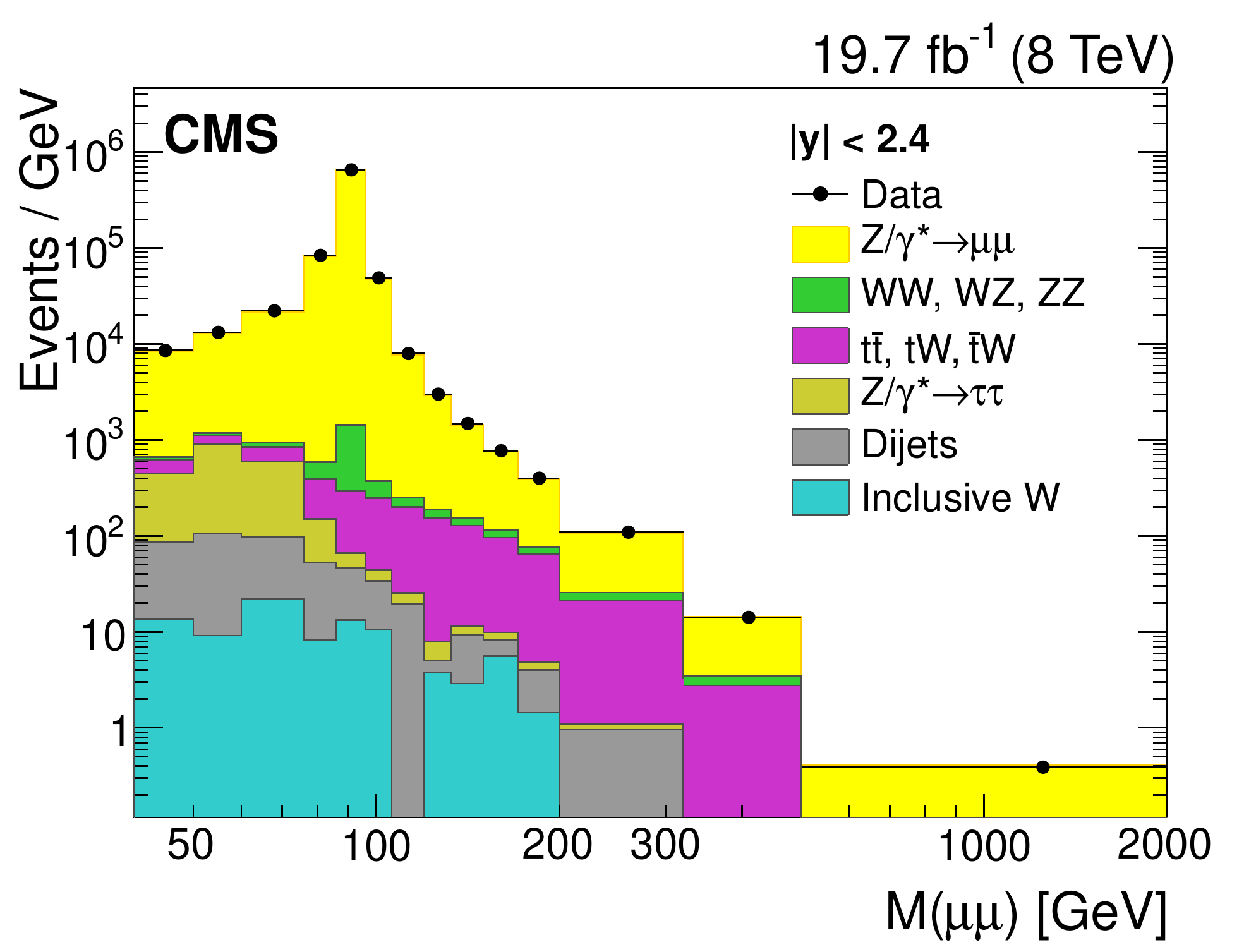}
\caption{\label{AFB1}Invariant mass distribution for dimuon events, rapidity up to 2.4 \cite{ref41}.}
\end{minipage}\hspace{2pc}%
\begin{minipage}{14pc}
\includegraphics[width=14pc]{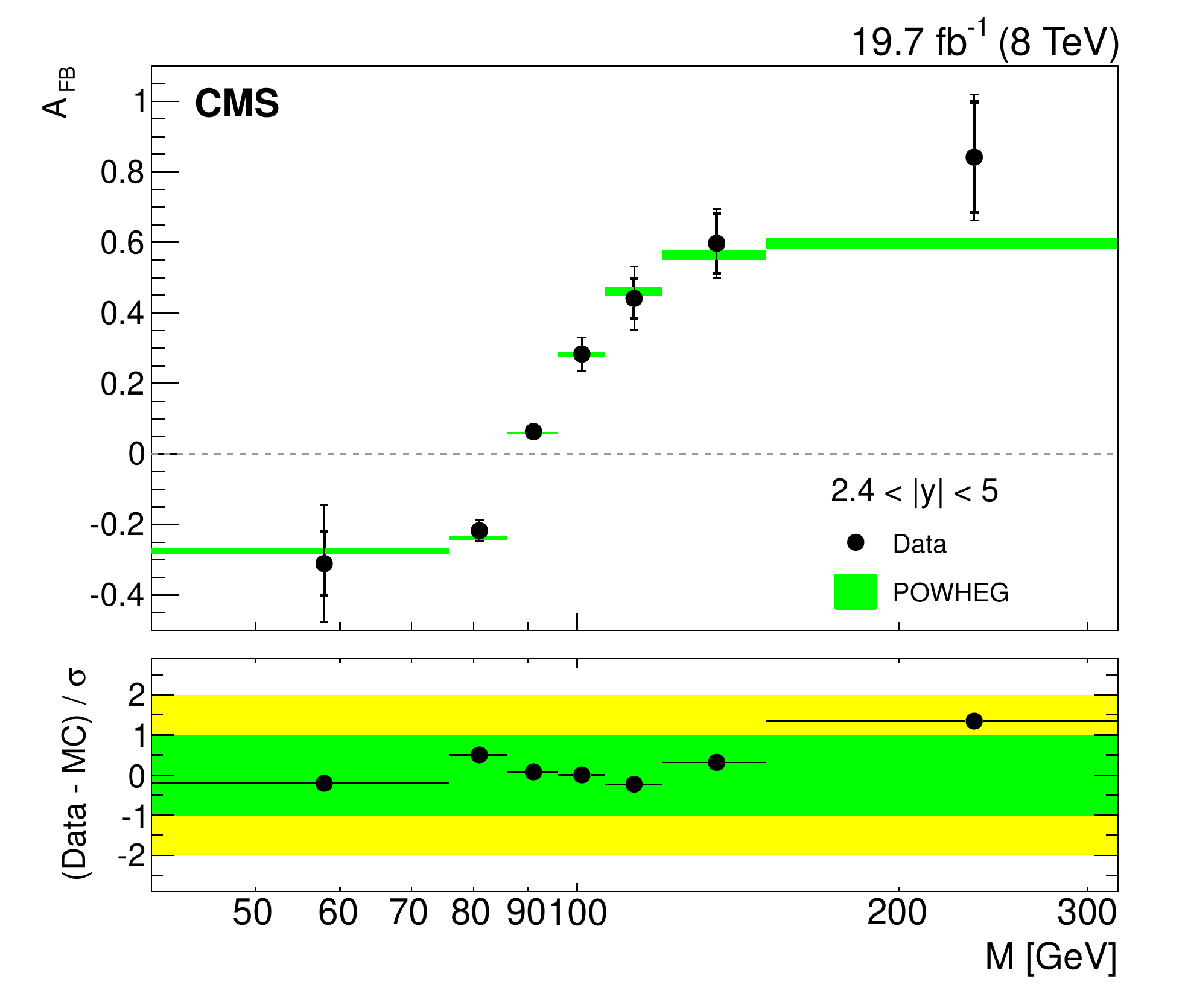}
\caption{\label{AFB2}The unfolded $A_{FB}$ distribution for the forward rapidity region \cite{ref41}.}
\end{minipage} 
\end{figure}

The forward-backward asymmetry $A_{FB}$ arises from the vector and axial-vector couplings of the electroweak bosons to fermions and it is measured \cite{ref41} for dilepton masses between 40 GeV and 2 TeV and dilepton rapidity up to 5. The invariant mass distribution for the dimuon events in shown in Fig.~\ref{AFB1}. The events are assigned either to "forward" or "backward" region based on the positive or negative value of $cos\Theta^{*}$, where $\Theta^{*}$ represents emission angle of the negatively charged lepton relative to the quark momentum in the rest frame of the dilepton system. Corrections were applied for detector resolution, acceptance and final state radiation. The effective weak mixing angle of $sin^{2}\Theta_{lept}^{eff}=0.2312$ is used for comparison with the prediction from POWHEG. For all rapidity regions the measured $A_{FB}$ values are in a good agreement with this prediction. The unfolded $A_{FB}$ distribution in the forward rapidity region is shown in Fig.~\ref{AFB2}.

\section{W boson differential cross section and charge asymmetry}

\begin{figure}[htb]
\begin{minipage}{14pc}
\includegraphics[width=14pc]{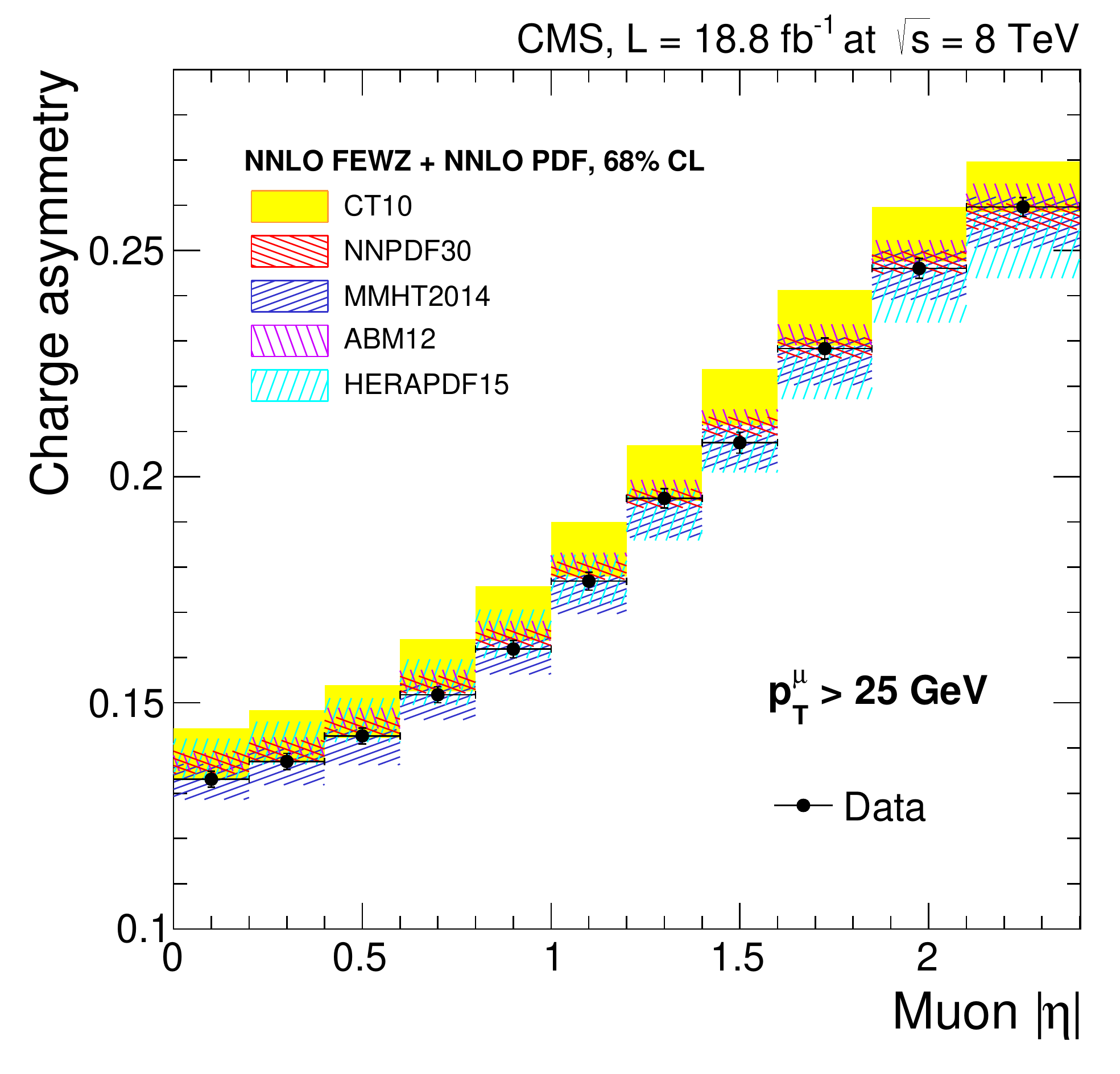}
\caption{\label{Wch1}Comparison of asymmetries to NNLO predictions calculated using FEWZ3.1 generator interface with different PDF sets \cite{ref51}.}
\end{minipage}\hspace{2pc}%
\begin{minipage}{14pc}
\includegraphics[width=14pc]{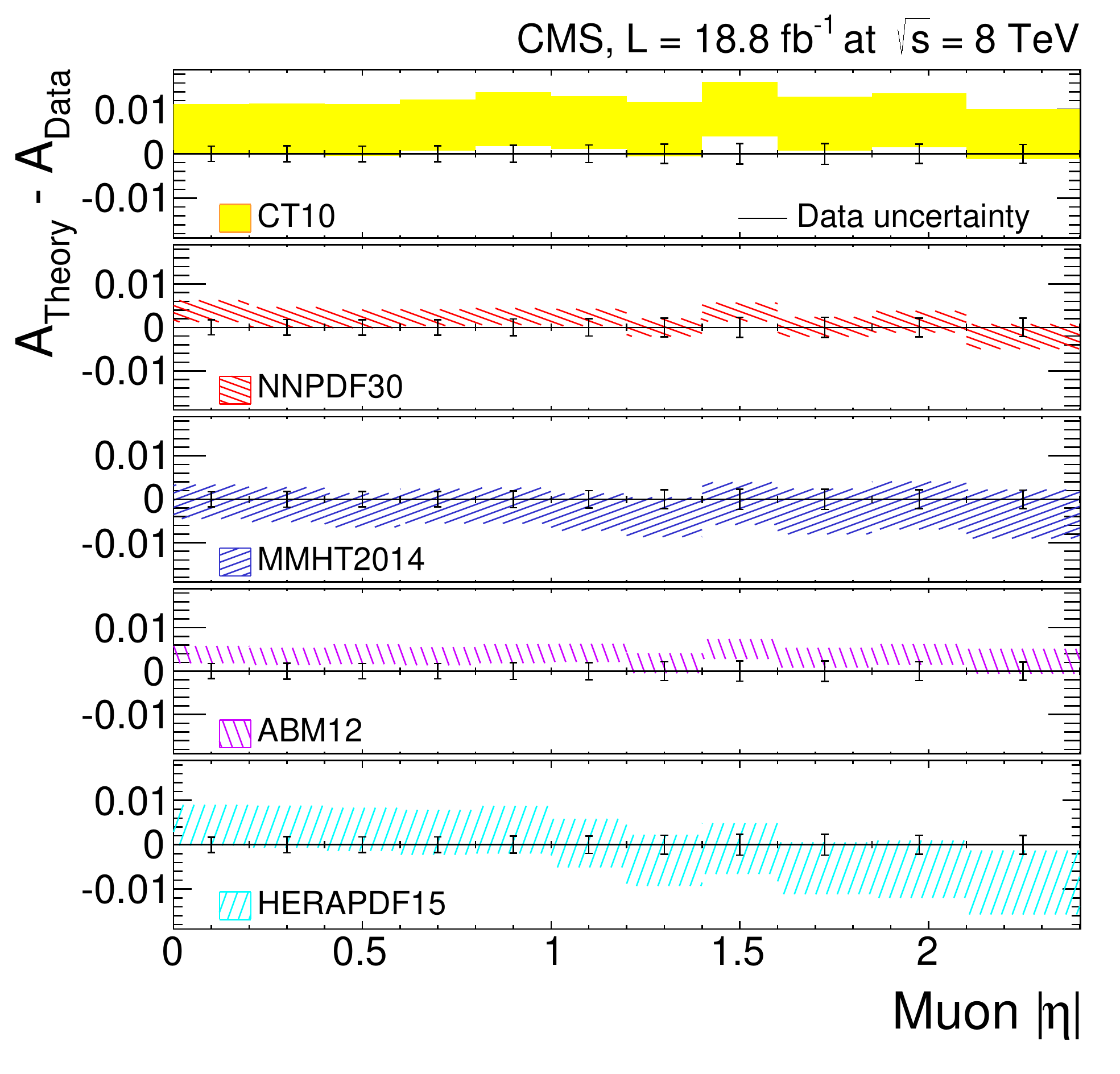}
\caption{\label{Wch2}Differences between the theoretical predictions and the measured asymmetries \cite{ref51}.}
\end{minipage} 
\end{figure}

The differential cross section is measured \cite{ref51} as a function of the muon pseudorapidity and found to be well described by the prediction from FEWZ generator and several parton density functions (PDF). No electroweak corrections are included for this measurement. Measured charge asymmetry also agrees well with the tested PDF sets, as shown in Fig.~\ref{Wch1} and Fig.~\ref{Wch2}. This measurement is able to provide constrains on the proton PDFs. The results are incorporated into QCD NNLO analysis and combined with the deep inelastic scattering data recorded by the HERA experiment \cite{ref52}. A significant improvement in the accuracy of the valence quark distributions in the low range of Bjoerken variable, $0.001 < x < 0.1$, has been observed.

\section{Z boson inclusive and differential cross section}

\begin{figure}[htb]
\begin{minipage}{14pc}
\includegraphics[width=16pc]{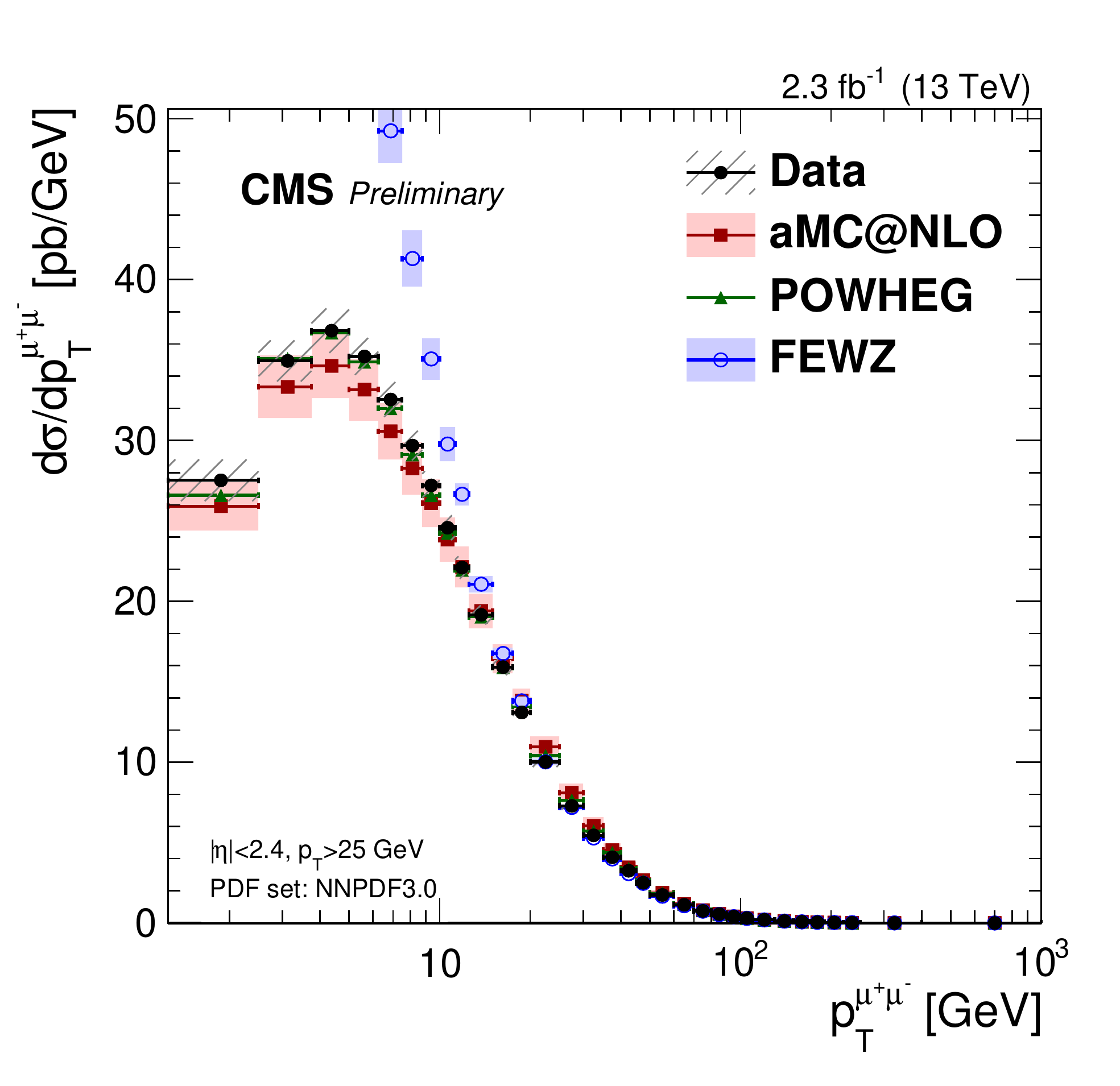}
\caption{\label{Zpt1}Transverse momentum distribution of the Z boson decaying to dimuons \cite{ref61}. }
\end{minipage}\hspace{2pc}%
\begin{minipage}{14pc}
\includegraphics[width=16pc]{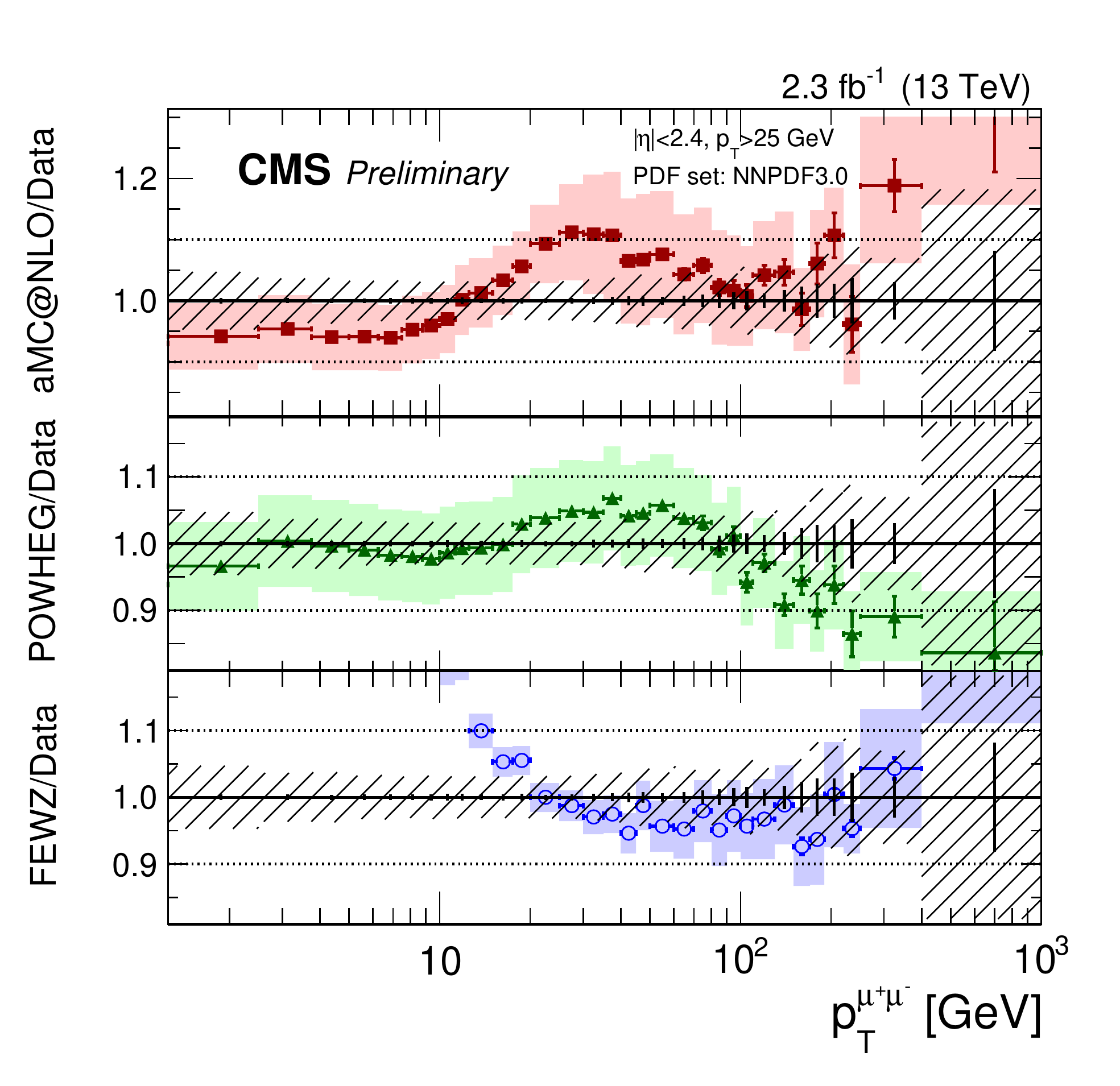}
\caption{\label{Zpt2}The MC prediction to data ratio for the dimuon transverse momentum \cite{ref61}. }
\end{minipage} 
\end{figure}

The inclusive and differential Z boson cross section produced via Drell-Yan process is measured \cite{ref61} as a function of kinematic variables. The transverse momentum distribution of the Z boson is a probe of the strong nuclear interaction. The total inclusive cross section times branching fraction is measured for the di-muon mass in the range of 60 to 120 GeV and found to have a value of $1870\pm2(stat)\pm35(syst)\pm51(lumi)pb$. This result is in a good agreement with the NNLO QCD and NLO electroweak calculations from FEWZ generator and various sets of PDFs used. The comparison of the measured differential cross section to MadGraph5aMC@NLO, POWHEG and FEWZ generators is shown in Fig.~\ref{Zpt1} and Fig.~\ref{Zpt2}. A good agreement with FEWZ is found, but a deviation at the low transverse momentum is observed due to absence of resummation. The MadGraph5 and POWHEG slightly over predict the distribution, within the uncertainties.

\section{W-like measurement of Z boson mass}

\begin{figure}[htb]
\begin{minipage}{14pc}
\includegraphics[width=14pc]{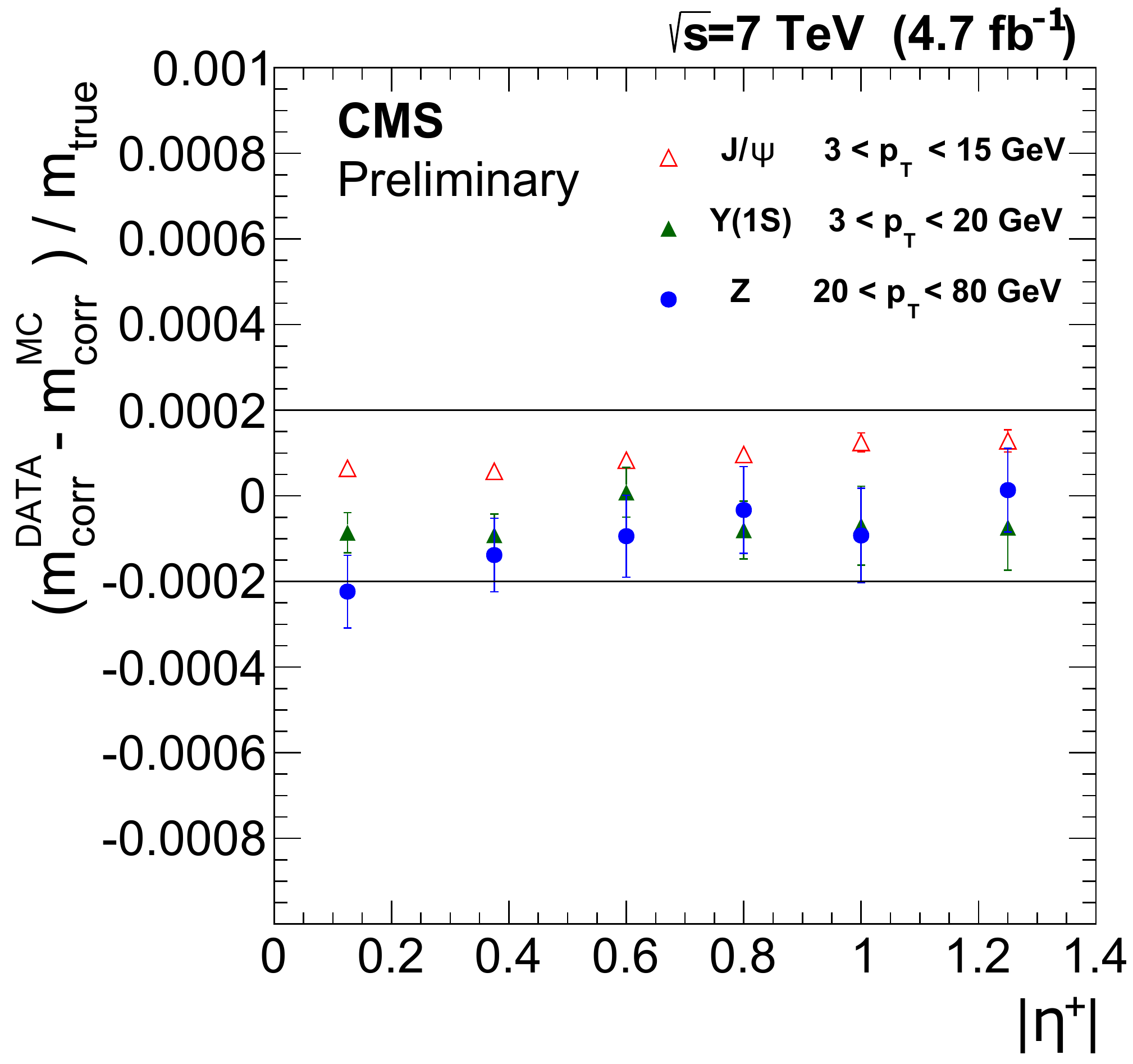}
\caption{\label{W1}Closure of the calibration of the relative scale (data with respect to MC) for $J/\Psi$,$\Upsilon(1S)$ and Z dimuons, as a function of transverse momentum of the positive muon \cite{ref71}.  }
\end{minipage}\hspace{2pc}%
\begin{minipage}{14pc}
\includegraphics[width=18pc]{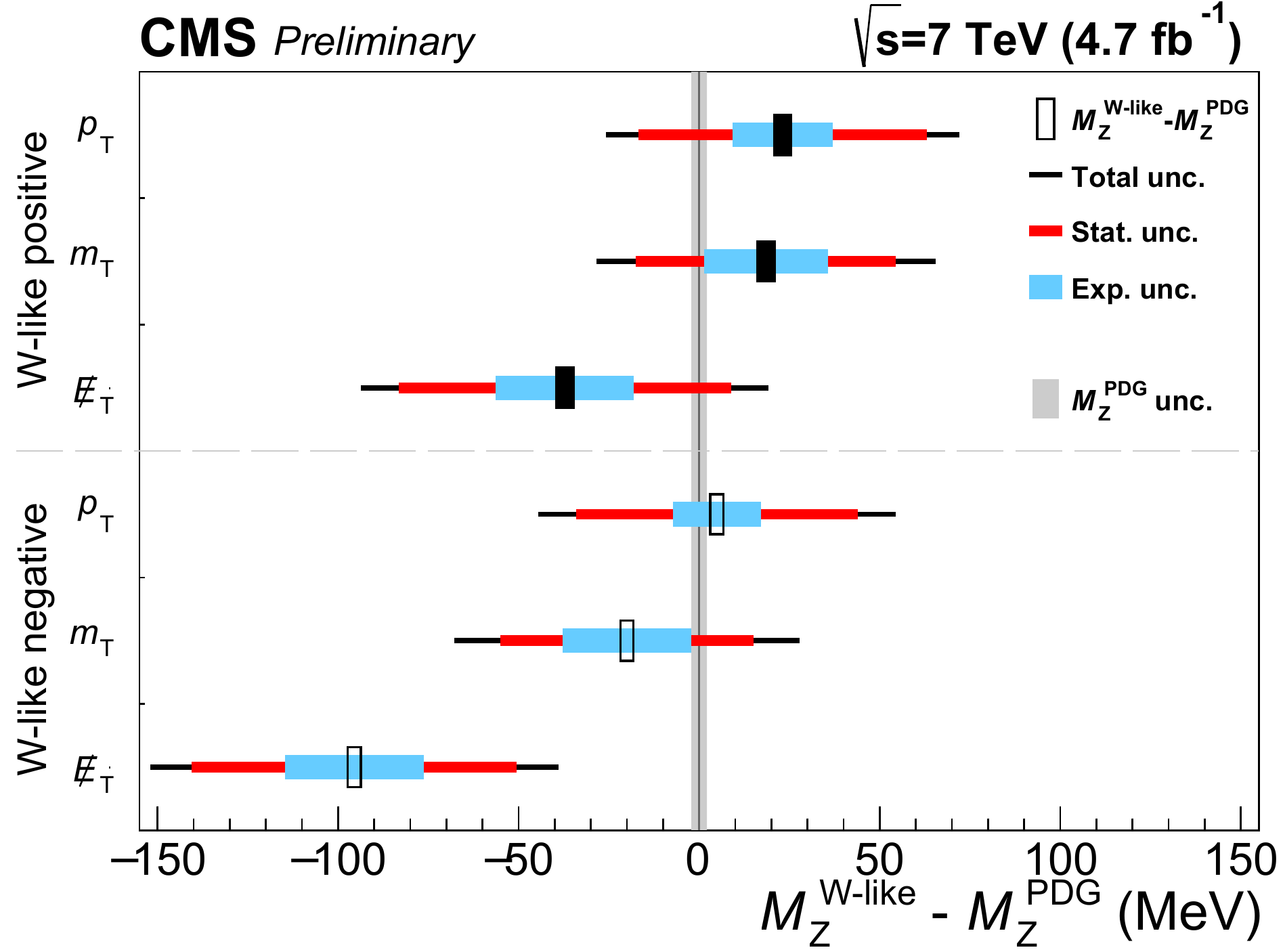}
\caption{\label{W2}Difference between the fitted mass and $M_{Z}^{PDG}$ obtained with each of the three observables \cite{ref71}.}
\end{minipage} 
\end{figure}

Precise measurement of the W, top and Higgs mass represent a crucial test of the SM. In this measurement \cite{ref71} the dimuon events from the Z boson decay were selected and one of the two muons removed to mimic the W boson decay topology. This is proof of principle and a test of the analysis methods for the future W mass measurement. The muon momentum was calibrated using the $J/\Psi$ and $Y(1s)$ dimuon events, as shown in Fig.~\ref{W1}. To reach an accurate control over the missing transverse energy, the hadronic recoil calibration is performed using dimuon events from Z boson decay. The results of the fits to data are shown in Fig.~\ref{W2} for the lepton transverse momentum, transverse mass and missing transverse energy, in the positive and negative "W-like" mass cases, where the experimental uncertainties on the recoil and the lepton are quoted separately from the others. The observable expected to provide the most precise measurement of the W mass, the transverse mass, provides a fitted mass that differs from the PDG value by 18 MeV or -20 MeV when using positive or negative samples, respectively. The future W mass measurement requires these calibration methods to be refined and the presented analysis methods to be extrapolated to the W mass case.

\section{Conclusions}

In the past years a lot of work has been done to understand the SM at the highest possible precision. The measurements clearly profited from the advances in theory calculations and MC generators. In this overview the differential cross sections and their ratios were compared to the NNLO predictions. The angular Z boson coefficients were measured and forward-backward asymmetry studied in the extended range of rapidity up to 5. The accuracy of valence quark distributions was improved with the measurement of the charge asymmetry of the W boson production. The proof of principle and validation of analysis methods for W mass measurement was established.

\Acknowledgements
I thank the CMS Collaboration for the work presented here and for the support in preparing this talk, the CERN accelerator division for the excellent operation of the LHC and all funding agencies that made this experiment possible.

\end{document}